# Thermodynamics and kinetics of vapor bubbles nucleation in one-component liquids

Nikolay V. Alekseechkin

Akhiezer Institute for Theoretical Physics, National Science Centre "Kharkov Institute of Physics and Technology", Akademicheskaya Street 1, Kharkov 61108, Ukraine

E-mail: n.alex@kipt.kharkov.ua

The multivariable theory of nucleation [J. Chem. Phys. **124**, 124512 (2006)] is applied to the problem of vapor bubbles formation in pure liquids. The presented self-consistent macroscopic theory of this process employs thermodynamics (classical, statistical and linear non-equilibrium), hydrodynamics and interfacial kinetics. As a result of thermodynamic study of the problem, the work of formation of a bubble is obtained and parameters of the critical bubble are determined. The variables $V$ (the bubble volume), $\rho$ (the vapor density), and $T$ (the vapor temperature) are shown to be natural for the given task. An equation for the dependence of surface tension on bubble state parameters is obtained. An algorithm of writing the equations of motion of a bubble in the space $\{V, \rho, T\}$ - equations for $\dot{V}$, $\dot{\rho}$, and $\dot{T}$ - is offered. This algorithm ensures symmetry of the matrix of kinetic coefficients. The equation for $\dot{T}$ written on the basis of this algorithm is shown to represent the first law of thermodynamics for a bubble. The negative eigenvalue of the motion equations which alongside with the work of the critical bubble formation determines the stationary nucleation rate of bubbles is obtained. Various kinetic limits are considered. One of the kinetic constraints leads to the fact that the nucleation cannot occur in the whole metastable region; it occurs only in some subregion of the latter. Zel'dovich' theory of cavitation is shown to be a limiting case of the theory presented. The limiting effects of various kinetic processes on the nucleation rate of bubbles are shown analytically. These are the inertial motion of a liquid as well as the processes of particles exchange and heat exchange between a bubble and surrounding liquid. The nucleation rate is shown to be determined by the slowest kinetic process at positive and moderately negative pressures in a liquid. The limiting effect vanishes at high negative pressures.

# I. INTRODUCTION

The kinetics of bubbles nucleation in a metastable liquid [1-9] (or the kinetics of boiling up) is one of the classical problems of the nucleation theory; its studying starts from the work of Doring and Volmer [1]. An important milestone in the development of the theory of bubbles formation is the classical work of Zel'dovich [2] in which the one-dimensional theory of cavitation in a liquid at high negative pressures is suggested (the presence of vapor in a bubble is neglected). A significant idea to use the macroscopic equations of motion of a nucleus in the space of its parameters for determining the kinetic coefficients of the nucleation theory (diffusivities in the Fokker-Planck equation) is offered in this paper. Hydrodynamic equations are employed for this purpose. Thereby it becomes possible to determine the limiting effects of various processes on nucleation: viscosity, heat conductivity and diffusion. The importance of this approach is obvious not only for cavitation, but also for other problems of nucleation, so it is universal. The use of this approach makes the nucleation theory consistent, i.e. it becomes fully *macroscopic*. The nucleation rate is determined only by macroscopic parameters of the mother phase both thermodynamic and hydrodynamic (the coefficients of viscosity, heat conductivity and diffusion) which can be measured.

At the same time, the approach which can be called "miscellaneous" is more common now in the nucleation theory. Macroscopic (thermodynamic) study is used for getting the work of nucleus formation, whereas the kinetic coefficients are obtained from microscopic consideration. Namely, kinetic processes on the interface are considered and the probabilities of forward and backward elementary processes are calculated. Historically the nucleation theory began to develop just by this way. However, this approach is limited; it is not applicable to all tasks and does not take into account properly the kinetic properties of the mother phase.

Returning to the boiling up of a superheated liquid at positive and negative pressures, it is clear that a multivariable theory of nucleation [10-14] is necessary to investigate this process in detail, i.e. to clarify the effect of the kinetic processes mentioned above on nucleation. The multivariable theory of nucleation of vapor bubbles presented in the given report uses the maximum number of variables – three, differently from earlier one-[2, 4] and two-variable [5] theories. This number would be equal to two according to the Gibbs phase rule if either the number of vapor particles or vapor density was constant; however, these are not the cases. As a consequence, all the limiting effects in this problem are taken into account properly: the inertial motion of a liquid, the evaporation-condensation processes, and the heat exchange between a bubble and surrounding liquid. The importance of thermal processes in calculating the nucleation rate of bubbles was noted in Ref. 3. The Einstein- Smoluchowski approach [14] is employed in the presented theory, i.e. the case of sufficiently viscous liquids is considered. However, the criterion of applicability of this approach involves the bubble critical size [9]; hence, it can be satisfied in a certain region of metastability for "not very viscous" liquids also.

The paper is organized as follows. A detailed thermodynamic consideration of the problem is carried out in Sec. II. The work of vapor bubble formation is obtained here. Further, parameters of the critical bubble and the most convenient variables for the given problem are determined. The dependence of surface tension on bubble state parameters is also considered and an equation for this dependence is obtained (an analogue of the Gibbs adsorption equation in the equilibrium theory). Sec. III focuses on the kinetic part of the problem, just getting the equations of motion of a bubble in the space of its variables. For this purpose, the algorithm of writing these equations is formulated and employed. Using these equations as well as the bubble formation work, the stationary nucleation rate of bubbles is calculated. Various kinetic limits are considered in Sec. IV. The limiting effects of different kinetic processes on the nucleation rate of bubbles are shown here.

## II. THERMODYNAMICS OF NUCLEATION

### A. Model and general equation for the work

We consider a vapor bubble in a liquid. The bubble volume $V$ is negligibly small in comparison with the volume $V_0$ of a liquid, so the bubble formation does not change the thermodynamic state parameters of the latter – the pressure $P_0$ and the temperature $T_0$. At the same time, a bubble is assumed to be a macroscopic subsystem, so we can apply the thermodynamic approach and the Fokker-Planck equation to it. Hereafter the quantities relating to a liquid will be provided by the subscript $0$, the quantities relating to a nucleus will be used without index; the critical nucleus parameters will be denoted by asterisk. In view of smallness of the bubble size, we can assume that the characteristic times of relaxation processes inside a bubble are sufficiently small (in comparison with other characteristic times of the problem), so the thermodynamic equilibrium takes place here at any time. The ambient phase (liquid) is in the state of metastable equilibrium. In other words, each of the coexisting phases is equilibrium, but there is no equilibrium between them; such equilibrium takes place only for the bubble of critical size. Nevertheless, considering near-critical bubbles, we can assume the deviations from the equilibrium of the whole system "nucleus + ambient phase" small and apply the classical thermodynamic approach to such quasiequilibrium system.

The main goal of thermodynamic consideration is to determine the work of nucleus formation. For this purpose, the expression for the minimum work $W$ done by a system (thermostat) on its subsystem [15, 16] has to be employed:

$$W = \Delta E - T_0 \Delta S + P_0 \Delta V \qquad (1)$$

The changes of energy, $\Delta E$, entropy, $\Delta S$, and volume, $\Delta V$, of the subsystem relate in our case to the region occupied by the new-phase nucleus which is a bubble. Since the state of the ambient phase does not change under nucleus formation, these quantities can be attributed to the whole system also. This general expression for $W$ is a consequence of the first and second laws of thermodynamics. It is employed in Ref. [15] for studying fluctuations. A nucleus is also a fluctuation, but heterophase one [17]. So, there exists the interface, as distinct from the case of a homophase fluctuation.

The work $W$ determines the change $\Delta S$ of the entropy of a system upon a fluctuation [15] or nucleus formation:

$$\Delta S = S_2 - S_1 = -\frac{W}{T_0}, \qquad (2)$$

where $S_1$ is the entropy of the system without a nucleus (the initial state), $S_2$ is the entropy of the system "nucleus + ambient phase". According to Einstein's formula, the probability of a fluctuation is proportional to $\exp(\Delta S / k)$, where $k$ is the Boltzmann constant. Therefore, the work (1) determines the equilibrium distribution function of nuclei

$$f_{eq}(\{x_i\}) = C_{eq} e^{-\frac{W(\{x_i\})}{kT_0}}, \qquad (3)$$

where $\{x_i\}$ is the set of variables describing a nucleus. In Ref. [13], the normalizing constant $C_{eq}$ has been determined for multivariable nucleation processes including binary nucleation.

The well known conditions of equilibrium in thermodynamics are obtained from the condition of entropy maximum $dS = 0$. However, this equation is only the necessary condition of an extremum, but not sufficient one. There is the state with $dS = 0$ which does not correspond to the entropy maximum. This is the case of nucleation, just the state "critical nucleus + ambient phase". Considering deviations from this state (denoted by asterisk), let us transform eq. (2) as follows:

$$\Delta S = S_2 - S_1 = -\frac{W_* + \Delta W}{T_0} = S_2^* - S_1 - \frac{\Delta W}{T_0},$$

from where

$$S_2 - S_2^* = -\frac{\Delta W}{T_0} = -\frac{W - W_*}{T_0} \qquad (4)$$

The equilibrium condition, $dS_2 = 0$, leads, in view of eq. (4), to

$$dW = 0 \tag{5}$$

It will be shown below that this equation determines the parameters of the critical nucleus.

The second differential $d^2S$ shows the type of an extremum. In the case of a usual (homophase) fluctuation, we have $d^2S_2 < 0$ which corresponds to the maximum of entropy for the state 1 (without a fluctuation or a nucleus); this is the stable equilibrium state. If we consider deviations from the state "critical nucleus + ambient phase", the quadratic form $d^2S_2$ is not of a definite sign, so it represents the saddle surface in the space $\{x_i\}$; the mentioned state itself corresponds to the saddle point [10-13]. Thus, the equilibrium of the critical nucleus with ambient phase is unstable; one of the variables $\{x_i\}$ (relating to the nucleus size) is unstable.

**B. Energy of the system "nucleus + ambient phase" and equations for surface tension**

From the two approaches to the description of interface phenomena- the finite-thickness layer method [18] and Gibbs' method [19] - the latter is used here in considering the thermodynamics of the heterogeneous system "nucleus + ambient phase" near the saddle point. The superficial (or excessive) quantities [19] relating to the interface between a nucleus and ambient phase are denoted by the subscript $\Sigma$; these are the energy $E_\Sigma$, the entropy $S_\Sigma$, and the number of particles $N_\Sigma$. The energy of the initial homogeneous system (without a nucleus) consisting of $N_{tot}$ particles and having the volume $V_1$ is [18]

$$E_1 = T_0 S_1 - P_0 V_1 + \mu_0 N_{tot}, \tag{6}$$

where $\mu_0$ is the chemical potential of the homogeneous ambient phase, $S_1$ is the entropy of the mentioned state 1.

The energy of the system "nucleus + ambient phase" can be represented as the sum of three parts: the energy of the homogeneous phase in the nucleus, $E = TS - PV + \mu N$, the energy of the homogeneous ambient phase, $E_0 = T_0 S_0 - P_0 V_0 + \mu_0 N_0$, and the superficial energy $E_\Sigma$:

$$E_2 = (TS - PV + \mu N) + (T_0 S_0 - P_0 V_0 + \mu_0 N_0) + E_\Sigma, \tag{7}$$

where $\mu$, $P$, $T$, $S$, and $N$ are respectively the chemical potential, the pressure, the temperature, the entropy and the particles number of the homogeneous phase in the nucleus (vapor in a bubble); the same quantities with the subscript 0 have the same meaning for the homogeneous ambient phase.

The entropy, $S_2$, and the volume, $V_2$, of the mentioned system are

$$S_2 = S + S_0 + S_\Sigma, \quad V_2 = V + V_0, \tag{8}$$

from where

$$\Delta S = S_2 - S_1 = S + S_0 + S_\Sigma - S_1, \quad \Delta V = V_2 - V_1 = V + V_0 - V_1 \tag{9}$$

Also

$$N + N_0 + N_\Sigma = N_{tot} \tag{10}$$

Calculating the difference $\Delta E = E_2 - E_1$, we use the equations $S_0 - S_1 = \Delta S - S - S_\Sigma$, $V_0 - V_1 = \Delta V - V$, and $N_0 - N_{tot} = -N - N_\Sigma$ which follows from eq's (9) and (10). Substituting $\Delta E$ into eq. (1), we find the following expression for the work:

$$W = (\mu - \mu_0)N + (T - T_0)S - (P - P_0)V + E_\Sigma - T_0 S_\Sigma - \mu_0 N_\Sigma \tag{11}$$

The following step is to get an expression for $E_\Sigma$. To this end, we have to extend properly the expression for the energy of a two-phase system "nucleus + ambient phase" being in equilibrium [18],

$$E_2 = T_0 S_2 - PV - P_0 V_0 + \sigma A + \mu_0 N_{tot}, \tag{12}$$

to a non-equilibrium (the mentioned above quasiequilibrium) case; this expression is presented in our designations. Here $\sigma$ is the surface tension, $A$ is the nucleus surface area; $T = T_0$ and $\mu = \mu_0$ in view of equilibrium.

The superficial entropy $S_\Sigma$, as an additive quantity, can be represented as the sum of two parts

$$S_\Sigma = \overline{S} + \overline{S}_0, \tag{13}$$

where $\overline{S}$ and $\overline{S}_0$ are the contributions to $S_\Sigma$ from the new and ambient phases respectively. The same is true for $N_\Sigma$:

$$N_\Sigma = \overline{N} + \overline{N}_0 \tag{14}$$

So, the direct generalization of eq. (12) looks as

$$E_2 = T(S + \overline{S}) + T_0(S_0 + \overline{S}_0) - PV - P_0 V_0 + \sigma A + \mu(N + \overline{N}) + \mu_0(N_0 + \overline{N}_0) \tag{15}$$

The separation of $S_\Sigma$ into two parts, eq. (13), solves the problem what temperature corresponds to $S_\Sigma$ in the equation for $E_2$. Each of the parts, $\overline{S}$ and $\overline{S}_0$, enters in eq. (15) with its own temperature, $T$ and $T_0$, respectively. The similar problem with $N_\Sigma$ and chemical potentials is solved by the separation yielded by eq. (14). So, eq. (15) is fully symmetric with respect to both the phases and seems a logical extension of eq. (12) to the quasiequilibrium case.

The desired quantity $E_\Sigma$ is determined now by comparison of eq.'s (7) and (15):

$$E_\Sigma = T\overline{S} + T_0 \overline{S}_0 + \sigma A + \mu \overline{N} + \mu_0 \overline{N}_0, \tag{16}$$

or, in equivalent form,

$$E_\Sigma = E_\Sigma^{(eq)} + (T - T_0)\overline{S} + (\mu - \mu_0)\overline{N}, \tag{17a}$$

$$E_\Sigma^{(eq)} = T_0 S_\Sigma + \sigma A + \mu_0 N_\Sigma, \tag{17b}$$

where $E_\Sigma^{(eq)}$ is the equilibrium value of $E_\Sigma$ in Gibbs' method [18, 19].

As is known, eq. (17b) together with the equation

$$dE_\Sigma = T_0 dS_\Sigma + \sigma dA + \mu_0 dN_\Sigma \tag{18}$$

lead to the Gibbs adsorption equation [18, 19]

$$Ad\sigma = -S_\Sigma dT_0 - N_\Sigma d\mu_0 \tag{19}$$

The generalized adsorption equation for the quasiequilibrium case is derived from eq.'s (17) and the following one:

$$dE_\Sigma = T_0 dS_\Sigma + \sigma dA + \mu_0 dN_\Sigma + (T - T_0)d\overline{S} + (\mu - \mu_0)d\overline{N} \tag{20}$$

It has the form

$$Ad\sigma = -S_\Sigma dT_0 - N_\Sigma d\mu_0 - \overline{S}d(T - T_0) - \overline{N}d(\mu - \mu_0) \tag{21}$$

In equilibrium, $T = T_0$ and $\mu = \mu_0$, it converts to eq. (19), as it must.

The properties of an interface have to depend on the properties of coexisting phases, so the surface tension is a function of thermodynamic parameters of both these phases. If these phases are in equilibrium, their thermodynamic parameters are not independent – they are connected with each other via the conditions of equilibrium. So, the surface tension in this case is a function of thermodynamic parameters of a one of the phases: $\sigma = \sigma_{eq}(P_*, T_*) = \sigma_{eq}(P_0, T_0)$; the equalities $T_* = T_0$ and $P_* = P_*(P_0, T_0)$ at the equilibrium point are shown below. In a non-equilibrium case, the surface tension depends on the new-phase parameters also: $\sigma = \sigma(P, T; P_0, T_0)$. Eq. (21) serves for determining such dependence (it should be noted that the finite-thickness layer method [18] is more convenient for this purpose). One of the assumptions of our model is the ambient phase parameters does not change upon the nucleus formation and further evolution, i.e. $T_0 = const$, $P_0 = const$, hence $\mu_0(P_0, T_0) = const$. So, eq. (21) is simplified and takes the form

$$(Ad\sigma)_{T_0, \mu_0} = -\overline{S}dT - \overline{N}d\mu \tag{22}$$

This equation determines the dependence $\sigma(P, T)$ of the surface tension on the nucleus parameters. Generally, this dependence has the form

$$\sigma(P, T) = \sigma_{eq}(P_0, T_0) + \overline{\sigma}(P, T) \tag{23}$$

which can be also regarded as an expansion of $\sigma(P, T)$ into a series in the vicinity of the equilibrium point (the saddle point):

$$A_*\bar{\sigma}(P,T) = -\bar{N}_*\upsilon(P-P_*) - (\bar{S}_* - \bar{N}_*s)(T-T_0) + \ldots, \tag{24}$$

where $\upsilon = \partial\mu/\partial P = V/N$ and $s = -\partial\mu/\partial T$ are the volume and the entropy per one particle in a nucleus.

Adding the Gibbs-Dugem equation for a homogeneous phase,

$$-SdT - Nd\mu + VdP = 0, \tag{25}$$

to eq. (22), we obtain one more representation of the latter:

$$(Ad\sigma)_{T_0,\mu_0} = -(S+\bar{S})dT + VdP - (N+\bar{N})d\mu \tag{26}$$

The quantities $(S+\bar{S})$ and $(N+\bar{N})$ are the *true* values of the nucleus entropy and number of particles.

### C. Work near the equilibrium point

Substituting eq.'s (17) into eq. (11), we obtain

$$W = (\mu-\mu_0)(N+\bar{N}) + (T-T_0)(S+\bar{S}) - (P-P_0)V + \sigma A \tag{27}$$

Calculating the differential of $W$ at constant $T_0$ and $P_0$, we employ eq.'s (22) and (25). As a result, one obtains

$$(dW)_{T_0,P_0} = (\mu-\mu_0)d(N+\bar{N}) + (T-T_0)d(S+\bar{S}) - (P-P_0)dV + \sigma dA \tag{28}$$

The condition of equilibrium, eq. (5), leads to the well known equations

$$T_* = T_0 \tag{29a}$$

$$\mu_* = \mu_0 \tag{29b}$$

$$P_* - P_0 = \frac{2\sigma}{R_*} \equiv P_L^*, \tag{29c}$$

where $R$ is the nucleus radius; $P_L^*$ is the Laplace pressure for the critical nucleus.

So, the dependence of surface tension on the nucleus state parameters does not change the classical conditions of equilibrium, as it must from the physical point of view.

Eq.'s (29) determine the parameters of the critical nucleus: the temperature, the vapor pressure or the density, and the critical radius. As is known [15, 16], the Thomson equation connecting the pressure $P_*$ of a saturated vapor in a bubble with the critical radius $R_*$ follows from eq.'s (29b, c) and has the form

$$P_* = P_\infty(T_0)e^{-\frac{2\upsilon_0\sigma}{kT_0R_*}}, \quad P_\infty(T_0) = Ce^{-\frac{q}{kT_0}}, \tag{30}$$

where $P_\infty(T_0)$ is the pressure of a saturated vapor at the plane interface; $\upsilon_0$ is the volume per one molecule in a liquid, $q$ is the heat of evaporation per one molecule, $C$ is the constant. Eq.'s (30) and (29c) are combined into a transcendental equation for the critical radius:

$$P_0 + \frac{2\sigma}{R_*} = P_\infty(T_0) e^{-\frac{2\upsilon_0 \sigma}{kT_0 R_*}} \tag{31}$$

Substitution of eq.'s (29) into eq. (27) leads to the well known Gibbs equation for the critical work

$$W_* = \frac{1}{3}\sigma A_* \tag{32}$$

Further step is to get the second differential of the work. The following relations are employed below:

$$S = sN, \quad V = \upsilon N, \quad A = gV^{2/3}, \quad g = 3^{2/3}(4\pi)^{1/3}, \tag{33}$$

and

$$\left[\sigma d^2 A - (P - P_0)d^2 V\right]_* = -\frac{2}{9} g\sigma V_*^{-4/3}(dV)^2, \tag{34}$$

Calculating the differential of eq. (28) at constant $T_0$, $P_0$, we obtain

$$(d^2 W)_{T_0,P_0} = d\mu d(N + \overline{N}) + dTd(S + \overline{S}) - dPdV + d\sigma dA + \left(\sigma d^2 A - (P - P_0)d^2 V\right) \tag{35}$$

In view of the equations $d\mu = -sdT + \upsilon dP$, $dS = sdN + Nds$, and $dV = \upsilon dN + Nd\upsilon$, eq. (35) is transformed as follows:

$$(d^2 W)_{T_0,P_0} = N(dTds - dPd\upsilon) + \left(\sigma d^2 A - (P - P_0)d^2 V\right) + \left[d\mu d\overline{N} + dTd\overline{S} + d\sigma dA\right] \tag{36}$$

Let us assume that the expression in square brackets in this equation is equal to zero and consider the meaning of this approximation. So, differentiating eq. (22), we have

$$d\mu d\overline{N} + dTd\overline{S} + d\sigma dA = -Ad^2\sigma - \overline{S}d^2 T - \overline{N}d^2\mu = 0, \tag{37}$$

from where

$$(Ad^2\sigma)_{T_0,P_0} = -\overline{S}d^2 T - \overline{N}d^2\mu \tag{38}$$

and

$$d\sigma = -\frac{d\overline{S}}{dA}dT - \frac{d\overline{N}}{dA}d\mu \tag{39}$$

Comparing eq.'s (39) and (22), we find that the considered approximation means

$$\frac{d\overline{S}}{dA} = \frac{\overline{S}}{A} \quad \text{and} \quad \frac{d\overline{N}}{dA} = \frac{\overline{N}}{A}, \tag{40a}$$

i.e.
$$\overline{S} = C_{\overline{S}}A, \quad \overline{N} = C_{\overline{N}}A, \tag{40b}$$

where $C_{\overline{N}}$ and $C_{\overline{S}}$ are the constants.

The proportionality of superficial quantities to the interface surface area is their natural property. The quantities $C_{\overline{S}}$ and $C_{\overline{N}}$ can be considered as constants at least in some vicinity of the equilibrium point. Starting from eq.'s (40b), we come to eq. (37). Thus, the given approximation seems to be physically plausible and the second differential at the equilibrium point has the form

$$(d^2W)_* = N_*[dTds - dPd\upsilon]_* - \frac{2}{9}g\sigma V_*^{-4/3}(dV)^2 \tag{41}$$

Here, as well as in eq.'s (29c)-(32), the surface tension $\sigma$ has its equilibrium value $\sigma_* \equiv \sigma_{eq}$.

So, in the vicinity of the equilibrium point [13]

$$W = W_* + \frac{1}{2}(d^2W)_* = W_* + \frac{1}{2}\sum_{i,k} h_{ik}(x_i - x_i^*)(x_k - x_k^*) \tag{42}$$

### D. Canonical variables

Using eq. (41), we can find the matrix **H** in eq. (42) for different sets of variables $\{x_i\}$ describing a vapor bubble.

**(1)** $\{x_i\} = (V, \rho, T)$, $\rho = N/V$ is the density of the homogeneous vapor phase.

The familiar thermodynamic relation

$$\left(\frac{\partial S}{\partial P}\right)_T = -\left(\frac{\partial V}{\partial T}\right)_P \tag{43}$$

can be employed for the determination of $ds(\rho, T)$. Together with the equation of state of vapor,

$$P(\rho, T) = \rho kT, \tag{44}$$

it yields $(\partial s/\partial \rho)_T = -k/\rho$, so

$$ds(\rho, T) = -\frac{k}{\rho}d\rho + \frac{c_V}{T}dT, \tag{45}$$

where $c_V$ is the heat capacity of vapor per one molecule at constant $V$. This equation also can be gotten from the equation for the entropy of an ideal gas [15] per one molecule:

$$s(\rho, T) = k\ln\frac{e}{\rho} + c_V \ln T + const \tag{46}$$

From the equation of state, eq. (44),

$$dP(\rho,T) = kTd\rho + k\rho dT \qquad (47)$$

Substituting eq.'s (45) and (47) into eq. (41) and taking into account $d\upsilon = d(1/\rho) = -d\rho/\rho^2$, we obtain the matrix $\mathbf{H}$:

$$\mathbf{H} = \begin{pmatrix} -\dfrac{2}{9}g\sigma V_*^{-\frac{4}{3}} & 0 & 0 \\ 0 & \dfrac{kT_0 V_*}{\rho_*} & 0 \\ 0 & 0 & \dfrac{c_V \rho_* V_*}{T_0} \end{pmatrix} \qquad (48)$$

The matrix $\mathbf{H}$ has a canonical form in the variables $(V,\rho,T)$, so these variables are canonical for the system considered. The variable $V$ is a natural unstable variable for a bubble, the quadratic form $N_*[dTds - dPd\upsilon]_*$ is positive definite. The hypersurface represented by the quadratic form $H(\{x_i\})$ is a saddle one. Note that the similar expression, $[\Delta T\Delta S - \Delta P\Delta V]$, is employed in Ref. [15] for calculating the fluctuations of thermodynamic quantities. Hence, the fluctuations of stable variables for the critical nucleus are the same as in the theory of fluctuations [15]. They can be found from eq. (48) according to the relation [13] $\langle(\Delta x_i)^2\rangle = kT_0/h_{ii}$. Thus, for the relative fluctuations of $\rho$ and $T$, one obtains

$$\frac{\sqrt{\langle(\Delta\rho)^2\rangle}}{\rho_*} = \frac{1}{\sqrt{N_*}}, \qquad \frac{\sqrt{\langle(\Delta T)^2\rangle}}{T_0} = \sqrt{\frac{k}{c_V}}\frac{1}{\sqrt{N_*}} \qquad (49)$$

From the physical point of view, the nucleation rate has not to depend on the choice of variables. Therefore, it is of interest to calculate the matrix $\mathbf{H}$ for other sets of variables.

**(2)** $\{x_i\} = (V, N, T)$

The following equations are employed:

$$ds(V,N,T) = \frac{k}{V}dV - \frac{k}{N}dN + \frac{c_V}{T}dT \qquad (50a)$$

$$dP(V,N,T) = -\frac{NkT}{V^2}dV + \frac{kT}{V}dN + \frac{kN}{V}dT \qquad (50b)$$

$$d\upsilon = \frac{dV}{N} - \frac{V}{N^2}dN \qquad (50c)$$

As a result, eq. (41) yields

$$\mathbf{H}_{(V,N,T)} = \begin{pmatrix} \left[\dfrac{N_* k T_0}{V_*^2} - \dfrac{2}{9} g\sigma V_*^{-\frac{4}{3}}\right] & -\dfrac{kT_0}{V_*} & 0 \\ -\dfrac{kT_0}{V_*} & \dfrac{kT_0}{N_*} & 0 \\ 0 & 0 & \dfrac{c_V N_*}{T_0} \end{pmatrix} \qquad (51)$$

**(3)** $\{x_i\} = (V, P, T)$

The following equations are employed:

$$ds(P,T) = -\frac{k}{P} dP + \frac{c_P}{T} dT \qquad (52\text{a})$$

$$d\upsilon = \frac{k}{P} dT - \frac{kT}{P^2} dP, \qquad (52\text{b})$$

the latter is obtained from the state equation in the form $\upsilon = kT/P$.

The matrix $\mathbf{H}_{(V,P,T)}$ is

$$\mathbf{H}_{(V,P,T)} = \begin{pmatrix} -\dfrac{2}{9} g\sigma V_*^{-\frac{4}{3}} & 0 & 0 \\ 0 & \dfrac{V_*}{P_*} & -\dfrac{V_*}{T_0} \\ 0 & -\dfrac{V_*}{T_0} & \dfrac{c_p P_* V_*}{kT_0^2} \end{pmatrix} \qquad (53)$$

The set $(V,P,T)$, as well as the set $(V,\rho,T)$, contains one unstable (extensive) variable and two stable (intensive) variables. However, the matrix $\mathbf{H}_{(V,P,T)}$ is not canonical, differently from the matrix $\mathbf{H}$, since the variables $P$ and $T$ are not independent - $P$ is the function of $T$ according to the state equation (44).

The set $(V,N,T)$ contains two extensive variables, $V$ and $N$. Both these variables are unstable. The transformation $N \to \rho = N/V$ retains only one unstable variable $V$, the variable $\rho$ is stable. Here the similarity with binary nucleation [10] takes place, where a nucleus is described by the numbers $n_1$ and $n_2$ of monomers of both species. Both these variables are physically equivalent; they are unstable. The transformation $(n_1, n_2) \to (n, c)$, where $n = n_1 + n_2$ and $c = n_2/n$, breaks this symmetry – the variable $n$ is unstable, the variable $c$ is stable [13, 20].

### III. KINETICS OF NUCLEATION

**A. Equations of motion of a nucleus and symmetry of kinetic coefficients**

The equations of motion of a nucleus in the space $\{x_i\}$ in the vicinity of the saddle point have the form [13]

$$\dot{x}_i = -z_{ik}(x_k - x_k^*), \quad \mathbf{Z} = \mathbf{DH}/kT, \tag{54}$$

where $\mathbf{D}$ is the matrix of diffusivities in the Fokker-Planck equation. The velocities $\dot{x}_i$ on definition (as they appear in the Fokker-Planck equation, $\dot{x}_i = \lim_{\Delta t \to 0}(\langle \Delta x_i \rangle_{\Delta t}/\Delta t)$) are macroscopic, i.e. $\dot{x}_i$ is the rate of change of the mean value $\langle x_i \rangle$ [2].

Our aim is to determine the negative eigenvalue $\kappa_1$ of the matrix $\mathbf{Z}$ which appears in the expression for the steady state nucleation rate [11-13]. The matrix $\mathbf{D}$ is shown [13, 21, 22] to be symmetric in nucleation processes, in accordance with Onsager's reciprocal relations. Hydrodynamic and phenomenological equations will be used here for deriving the matrix $\mathbf{Z}$. The question arises: whether they will provide symmetry of the matrix $\mathbf{D}$? To answer this question let us consider at first a general algorithm of writing the motion equations (54) which takes into account the symmetry conditions. Then we shall consider how the given algorithm agrees with the use of hydrodynamic and phenomenological equations.

Let the variable $x_1$ be unstable (the volume $V$ of a nucleus); the remaining variables $x_i$, $i > 1$, are stable. The incrementing $\delta x_i$ of a stable variable, $i > 1$, in the elementary act consists of the two parts – the regular one which is proportional to the incrementing $\delta x_1$ of the unstable variable and the fluctuating one [13, 21, 22], $\delta \tilde{x}_i$:

$$\delta x_i = a_i \delta x_1 + \delta \tilde{x}_i \tag{55}$$

In accordance with this fact, eq.'s (54) for stable variables can be represented in the following form:

$$\dot{x}_i = a_i \dot{x}_1 + (\dot{x}_i)_{x_1}, \quad i > 1, \tag{56}$$

where the addend $(\dot{x}_i)_{x_1}$ means that the derivative $\dot{x}_i$ is calculated at constant value of $x_1$. Hence, $(\dot{x}_i)_{x_1}$ is proportional to stable variables only:

$$(\dot{x}_i)_{x_1} = -\lambda_{ik}(x_k - x_k^*), \quad i,k > 1 \tag{57}$$

These equations are nothing but the familiar equations of linear non-equilibrium thermodynamics [15].

As is evident from the foregoing thermodynamic treatment, the number of variables in our problem is equal to three, correspondingly, the system (57) consists of two equations for two stable variables $x_2$ and $x_3$. Comparing eq.'s (56) and (57) with eq. (54), it is easy to get expressions for $a_i$ as well as relations between elements of the matrices $\Lambda$ and $Z$:

$$a_2 = \frac{z_{21}}{z_{11}}, \quad a_3 = \frac{z_{31}}{z_{11}} \tag{58a}$$

$$\lambda_{22} = \frac{\overline{z}_{33}}{z_{11}}, \quad \lambda_{23} = -\frac{\overline{z}_{32}}{z_{11}}, \quad \lambda_{32} = -\frac{\overline{z}_{23}}{z_{11}}, \quad \lambda_{33} = \frac{\overline{z}_{22}}{z_{11}}, \tag{58b}$$

where the line denotes an algebraic adjunct to the corresponding matrix element. Eq.'s (58b) can be also presented in equivalent form as follows:

$$z_{22} = \lambda_{22} + a_2 z_{12}, \quad z_{23} = \lambda_{23} + a_2 z_{13}, \quad z_{32} = \lambda_{32} + a_3 z_{12}, \quad z_{33} = \lambda_{33} + a_3 z_{13} \tag{58c}$$

As noted above, the matrix $\mathbf{D} = kT\,\mathbf{ZH}^{-1}$ has to be symmetric. The matrix $\mathbf{H}$ in our problem is canonical, eq. (48). The conditions of symmetry of the matrix $\mathbf{D}$ yield the following relations:

$$a_2 z_{11} h_{11}^{-1} = z_{12} h_{22}^{-1} \tag{59a}$$

$$a_3 z_{11} h_{11}^{-1} = z_{13} h_{33}^{-1} \tag{59b}$$

$$z_{32} h_{22}^{-1} = z_{23} h_{33}^{-1}, \tag{59c}$$

where $h_{ii}^{-1}$ is an element of the matrix $\mathbf{H}^{-1}$; $h_{ii}^{-1} = 1/h_{ii}$ in our case.

The similar matrix $\mathbf{D}_{st} = kT\mathbf{\Lambda H}_{st}^{-1}$ for the stable variables $x_2$ and $x_3$ is also symmetric according to Onsager's reciprocal relations; here

$$\mathbf{H}_{st}^{-1} = \begin{pmatrix} h_{22}^{-1} & 0 \\ 0 & h_{33}^{-1} \end{pmatrix} \tag{60}$$

From here

$$\lambda_{32} h_{22}^{-1} = \lambda_{23} h_{33}^{-1} \tag{61}$$

It is easy to show with the help of eq.'s (58c), (59a) and (59b) that eq.'s (59c) and (61) are identical.

Now the algorithm of writing eq.'s (54) based on the conditions of symmetry of kinetic coefficients can be formulated.

(1) Eq. (54) for $\dot{x}_1$ (the unstable variable derivative) is fully arbitrary, i.e. the matrix elements $z_{11}$, $z_{12}$ and $z_{13}$ are arbitrary.

(2) Eq. (57) for $\dot{x}_2$ is also arbitrary, i.e. the elements $\lambda_{22}$ and $\lambda_{23}$ are arbitrary.

(3) Only the summand proportional to $x_3$, i.e. the element $\lambda_{33}$, is arbitrary in eq. (57) for $\dot{x}_3$.

(4) The terms in eq.'s (56) which are not arbitrary, just the coefficients $a_2$, $a_3$ and $\lambda_{32}$, are determined from the conditions of symmetry, eq.'s (59) and (61). Then the remaining elements of the matrix $Z$ are determined from eq.'s (58a) and (58c).

Obviously, eq.'s (54) written arbitrarily, in general, do not lead to a symmetric matrix $\mathbf{D}$. As an example, the non-symmetric matrix $\mathbf{D}$ obtained in Ref. [5] can be presented. Only equations written in accordance with the given algorithm ensure symmetry of this matrix. After deriving the motion equations by this way, it is necessary to ascertain their physical meaning and to answer the question whether they are adequate to the considered process. Answers to these questions are given below by the example of the studied process of vapor bubbles formation.

Equations similar to eq.'s (59) and (61) as well as the algorithm of writing eq.'s (54) take place in the case of arbitrary (non-canonical) matrix $\mathbf{H}$ also. However, they are not placed here for brevity.

### B. Equations of motion of a vapor bubble

Eq.'s (54), (56) and (57) for a vapor bubble in the variables $(V, \rho, T)$ have the following explicit forms:

$$\begin{cases} \dot{V} = -z_{VV}(V - V_*) - z_{V\rho}(\rho - \rho_*) - z_{VT}(T - T_0) \\ \dot{\rho} = a_\rho \dot{V} + (\dot{\rho})_V \\ \dot{T} = a_T \dot{V} + (\dot{T})_V \end{cases} \quad (62a)$$

$$\begin{cases} (\dot{\rho})_V = -\lambda_{\rho\rho}(\rho - \rho_*) - \lambda_{\rho T}(T - T_0) \\ (\dot{T})_V = -\lambda_{T\rho}(\rho - \rho_*) - \lambda_{TT}(T - T_0) \end{cases} \quad (62b)$$

#### 1. Equation for bubble volume

The dynamics of a spherical cavity in an inviscid and incompressible liquid is described by the Rayleigh [23] equation

$$\rho_0 \left[ R\ddot{R} + \frac{3}{2}\dot{R}^2 \right] = P_R - P_0, \quad (63)$$

where $R$ is the cavity radius, $P_R$ is the pressure in a liquid at the cavity boundary ($P_0$ is the pressure far from the cavity), $\rho_0$ is the mass density of a liquid.

Rederivation of this equation for a viscous liquid taking into account the viscosity terms both in the Navier-Stokes equation and boundary conditions yields [4, 5]

$$\rho_0 R\ddot{R} + \frac{3}{2}\rho_0 \dot{R}^2 + 4\eta_0 \frac{\dot{R}}{R} = P - P_L - P_0, \quad (64)$$

where $\eta_0$ is the viscosity of a liquid, $P$ is the pressure of vapor in the bubble.

In terms of a volume, eq. (64) has the form

$$\ddot{V} = \frac{\dot{V}^2}{6V} - \frac{4\eta_0}{b\rho_0}\frac{\dot{V}}{V^{2/3}} + \frac{3}{b\rho_0}V^{1/3}\left(P - \frac{2\sigma}{\sqrt{b}V^{1/3}} - P_0\right), \quad b = (3/4\pi)^{2/3} \qquad (65)$$

In this equation, we neglect by the term proportional to $\dot{V}^2$, since the motion equations are linear in the vicinity of the saddle point, and put $\ddot{V} = 0$, since the case of high viscosity is considered. As a result, we have the following equation:

$$\dot{V} = 3\xi V\left(P - \frac{2\sigma}{\sqrt{b}V^{1/3}} - P_0\right), \quad \xi \equiv 1/4\eta_0 \qquad (66)$$

Substituting here $P = \rho kT$ and expanding the right side near the saddle point up to linear terms in accordance with eq. (54), we find

$$z_{VV} = -\xi P_L^*, \quad z_{V\rho} = -3\xi V_* kT_0, \quad z_{VT} = -3\xi kV_*\rho_* = -3\xi kN_* \qquad (67)$$

## 2. Equation for vapor density

According to the definition of density, $\rho = N/V$, its derivative is

$$\dot{\rho} = -\frac{\rho}{V}\dot{V} + \frac{1}{V}\dot{N} \qquad (68)$$

Equation for $\dot{N}$ [4, 5],

$$\dot{N} = \pi\beta u R^2\left[P_{eq}(T,R) - P\right]/kT = \pi\beta u R^2\left[\rho_{eq}(T,R) - \rho\right], \qquad (69)$$

is the difference of fluxes of evaporation, $\sim P_{eq}$, and condensation, $\sim P$. The flux of evaporation is assumed to be the same as in the equilibrium state for given $T$ and $R$ ($P = P_{eq}(T,R)$), when both the fluxes are equal to each other in accordance with the detailed balancing principle [13]; $P_{eq}(T,R)$ is given by the Thomson equation (30) with arbitrary $T$ and $R$. Here $u = \sqrt{8kT/\pi m}$ is the mean thermal velocity of vapor molecules, $\beta$ is the condensation coefficient, and

$$\rho_{eq}(T,R) = \frac{C}{kT}e^{-\frac{q}{kT}\left(1+\frac{2\upsilon_0\sigma}{qR}\right)}, \qquad (70)$$

The first summand in eq. (68) is proportional to $\dot{V}$, as in general equation (62a). In other words, eq. (68) initially has the form of eq. (62a). On the other hand, the form of the factor at $\dot{V}$ (the factor $a_\rho$) is dictated by one of the symmetry conditions, eq. (59a) (see point **4** of the algorithm):

$$a_\rho = \frac{z_{V\rho}}{z_{VV}}\frac{h_{VV}}{h_{\rho\rho}} \qquad (71a)$$

Substituting here expressions (67) for $z_{VV}$ and $z_{V\rho}$ as well as expressions for $h_{VV}$ and $h_{\rho\rho}$, eq. (48), we find

$$a_\rho = -\frac{\rho_*}{V_*} \tag{71b}$$

which is in full agreement with eq. (68) written near the saddle point.

So, the symmetry conditions do not change the natural form of the equation for $\dot\rho$. From eq.'s (68) and (62a),

$$(\dot\rho)_V = \frac{1}{V}(\dot N)_V \tag{72}$$

Expanding $(\dot N)_V$, eq. (69), in $\rho$ and $T$ near the saddle point up to linear terms, we determine the elements $\lambda_{\rho\rho}$ and $\lambda_{\rho T}$ in the first of eq.'s (62b):

$$\lambda_{\rho\rho} = \frac{3}{4}\frac{\beta u(T_0)}{R_*}, \quad \lambda_{\rho T} = -\lambda_{\rho\rho}\frac{\rho_*}{T_0}\tilde q, \quad \tilde q \equiv \frac{q - kT_0 + 2\upsilon_0\sigma/R_*}{kT_0} \tag{73}$$

So, the desired equation for $\dot\rho$ is

$$\dot\rho = -\frac{\rho_*}{V_*}\dot V - \lambda_{\rho\rho}(\rho - \rho_*) + \lambda_{\rho\rho}\frac{\rho_*}{T_0}\tilde q(T - T_0) \tag{74}$$

with $\lambda_{\rho\rho}$ given by eq. (73).

### 3. Equation for vapor temperature

According to the above algorithm, the coefficients $a_T$ and $\lambda_{T\rho}$ in eq.'s (62a) and (62b) are determined from symmetry conditions (59b) and (61):

$$a_T = \frac{z_{VT}}{z_{VV}}\frac{h_{VV}}{h_{TT}} = -\frac{kT_0}{c_V V_*} = -\frac{P_*}{C_V^*}, \tag{75}$$

$$\lambda_{T\rho} = \lambda_{\rho T}\frac{h_{\rho\rho}}{h_{TT}} = -\lambda_{\rho\rho}\frac{kT_0}{c_V \rho_*}\tilde q, \tag{76}$$

where $C_V^* = c_V N_*$ is the heat capacity of the critical bubble.

So, the equation for $\dot T$ is

$$\dot T = -\frac{P_*}{C_V^*}\dot V - \frac{1}{C_V^*}\left[q - kT_0 + \frac{2\upsilon_0\sigma}{R_*}\right](\dot N)_V - \lambda_{TT}(T - T_0), \tag{77a}$$

or, in equivalent form,

$$dE = C_V^* dT + \left[q - kT_0 + \frac{2\upsilon_0\sigma}{R_*}\right]dN = -P_*dV + dQ, \tag{77b}$$

where

$$dQ = -C_V^* \lambda_{TT}(T - T_0)dt = -4\pi R_*^2 \alpha(T - T_0)dt \tag{77c}$$

is the heat received by the bubble from the ambient liquid ($Q > 0$, if $T < T_0$) in the time interval $dt$. The heat exchange occurs according to Newton's law

$$j = \alpha(T - T_0), \tag{78}$$

where $j$ is the heat flux density, $\alpha$ is the coefficient of heat exchange between a vapor bubble and surrounding liquid.

The expression $\Delta\varepsilon = q - kT_0$ is the first law of thermodynamics for evaporating molecules (per one molecule): $q$ is the heat of evaporation received from the liquid, $P_*(\upsilon - \upsilon_0) \approx P_*\upsilon = kT_0$ is the work done upon evaporation. The quantity $2\upsilon_0 \sigma / R_* = \sigma \delta A_{(1)}$ is the work of the interface increase upon the evaporation of one molecule. The energy of an ideal gas is [15] $E = Nc_V T + N\varepsilon_0 = C_V T + const$. The constituent parts of a bubble are the vapor and the interface. The full change of the vapor bubble energy in left side of eq. (77b) contains the contributions from both these components: the change of the vapor energy due to the temperature change, the change of the bubble energy due to the addition of $dN$ evaporated interfacial molecules to the vapor and due to the increase of the interface area. In the right side of eq. (77b), we have the work done by the bubble during the change of its volume, $dV$, and the heat $dQ$ received by the bubble from surrounding liquid. So, equation (77a) for the temperature is nothing but the *first law of thermodynamics* for a vapor bubble. The second summand in this equation is the change of the bubble temperature (cooling) due to the evaporation of molecules from its boundary.

From eq. (77c),

$$\lambda_{TT} = \frac{4\pi R_*^2 \alpha}{C_V^*} = \frac{3\alpha}{c_V \rho_* R_*} \tag{79}$$

Finally, the equation for $\dot{T}$ is

$$\dot{T} = -\frac{P_*}{C_V^*}\dot{V} + \lambda_{\rho\rho}\frac{kT_0}{c_V \rho_*}\tilde{q}(\rho - \rho_*) - \frac{3\alpha}{c_V \rho_* R_*}(T - T_0) \tag{80}$$

So, the symmetry conditions for the kinetic coefficients not only do not impose some artificial restrictions on the motion equations, but they "help" us to write them correctly. Thereby, thermodynamics, hydrodynamics and interfacial kinetics are combined into a self-consistent theory of nucleation. The question on the applicability of Onsager's reciprocal relations in the nucleation theory in view of the presence of an unstable variable was discussed in Ref. [13]. These relations were shown to be a consequence of the detailed balancing principle. At the same time, the motion equations (54) involve this principle; they are obtained from the Fokker-Planck equation with the use of the equilibrium distribution function $f_{eq}(\{x_i\})$ [13]. The derivation of the equations for $\dot{\rho}$ and $\dot{T}$ demonstrated here also shows that the reciprocal relations are a consequence of the fundamentals of kinetics (the detailed balancing principle) and thermodynamics (the first law). This conclusion supports and complements the result of Ref. [13]. Thus, the reciprocal relations are validated by a phenomenological way, without resorting to a microscopic consideration.

### C. Relaxation of bubble temperature

It is necessary to express the heat exchange coefficient $\alpha$ via thermal characteristics of the liquid which are more definite and experimentally measured. For this purpose, the thermal problem of the bubble temperature relaxation has to be considered (without regard for the processes of evaporation-condensation). When the bubble temperature $T$ deviates from its equilibrium value $T_0$, it relaxes to the latter according to the equation of linear non-equilibrium thermodynamics [15]

$$\frac{dF(t)}{dt} = -\lambda_{TT} F(t), \quad F(t) \equiv T(t) - T_0, \tag{81a}$$

from where

$$F(t) = F(0) e^{-\lambda_{TT} t} \tag{81b}$$

It should be recalled that according to the one of the model assumptions, relaxation processes inside the bubble occur sufficiently fast, so the temperature inside the bubble is uniform at any time.

Thus, the problem of the temperature distribution, $T(r,t)$, around the sphere of radius $R$ and temperature $F(t)$ arises. Its solution is [24]

$$T(x,t) = \frac{Rx}{2(x+R)\sqrt{\pi\chi_0}} \int_0^t \frac{F(\tau)}{(t-\tau)^{3/2}} e^{-\frac{x^2}{4\chi_0(t-\tau)}} d\tau, \quad x \equiv r - R, \tag{82}$$

where $\chi_0$ is the thermal diffusivity of the liquid.

The heat flux density on the bubble boundary $r = R$ is

$$j(t) = -\theta_0 \frac{dT}{dx}\bigg|_{x=0} = \theta_0 \left\{ \frac{F(t)}{R} + \frac{F(0)}{\sqrt{\pi\chi_0 t}} + \frac{1}{\sqrt{\pi\chi_0}} \int_0^t \frac{dF}{d\tau} \frac{d\tau}{\sqrt{t-\tau}} \right\}, \tag{83a}$$

or, with regard for eq. (81b),

$$j(t) = \theta_0 \left\{ \frac{F(0) e^{-\lambda_{TT} t}}{R} + \frac{F(0)}{\sqrt{\pi\chi_0 t}} - \frac{2\sqrt{\lambda_{TT}}}{\sqrt{\pi\chi_0}} F(0) e^{-\lambda_{TT} t} \int_0^{\sqrt{\lambda_{TT} t}} e^{y^2} dy \right\}, \tag{83b}$$

where $\theta_0$ is the heat conductivity coefficient of the liquid.

The total heat $\Delta Q_\infty$ transferred across the sphere $r = R$ is

$$\Delta Q_\infty = 4\pi R^2 \int_0^\infty j(\tau)d\tau = 4\pi R^2 \frac{\theta_0 F(0)}{R\lambda_{TT}} \tag{84a}$$

On the other hand, according to Newton's heat-exchange law,

$$\Delta Q_\infty = 4\pi R^2 \alpha \int_0^\infty F(\tau)d\tau = 4\pi R^2 \alpha \frac{F(0)}{\lambda_{TT}} \tag{84b}$$

From comparing these expressions, the desired relation is obtained:

$$\alpha = \frac{\theta_0}{R} \tag{85}$$

### D. Nucleation rate of bubbles

The multivariable stationary nucleation rate calculated in Ref. [13] has the form

$$I = N_b \sqrt{\frac{kT_0}{2\pi} |h_{11}^{-1}|} |\kappa_1| e^{-\frac{W_*}{kT_0}}, \tag{86}$$

where $\kappa_1$ is the negative eigenvalue of the matrix $\mathbf{Z} = \mathbf{DH}/kT$ corresponding to the system of equations (54) (note that the matrix $\mathbf{Z} = \mathbf{DH}$ is employed in Ref. [13], so $\kappa_1$ differs by the multiplier $kT$ therein). $N_b$ is the normalizing constant of the one-dimensional equilibrium distribution function. If a nucleus is described by the number of monomers forming it (e.g. a drop in a vapor), then $N_b$ is the number of monomers of the parent phase in unite volume [13]. However, the determination of $N_b$ for bubbles is a separate problem. Its solution can be obtained within the framework of the hole theory of liquids [17] in view of the analogy between holes in liquids and vacancies in solids.

The equilibrium number of vacancies or holes in a liquid at given $T$ and $P$ is [17]

$$N_h = N_1 e^{-\frac{w_h(P,T)}{kT}}, \tag{87}$$

where $w_h(P,T)$ is the work of vacancy (the minimum-size hole in a liquid [17]) formation, $N_1 = 1/\upsilon_0$ is the number of atoms in a solid or monomers in a liquid in unite volume. Let $\omega_h$ be the volume of a vacancy or the minimum-size hole. The cluster of $n$ vacancies or minimum-size holes has the volume $V = n\omega_h$, so the equilibrium distribution function of multi-vacancy complexes or bubbles is

$$f_{eq}(V) = \frac{N_h}{\omega_h} e^{-\frac{W(V)}{kT}}, \tag{88}$$

where $W(V)$ is the work of formation of the bubble of volume $V$. The normalizing constant for bubbles, $N_b$, is obtained from eq.'s (87) and (88) as

$$N_b = \frac{1}{\upsilon_0 \omega_h} e^{-\frac{W_h(P_0, T_0)}{kT_0}} \tag{89}$$

Having the equations for $\dot{\rho}$ and $\dot{T}$, we can get the remaining elements of the matrix $\mathbf{Z}$ according to eq.'s (58a) and (58c):

$$\mathbf{Z} = \begin{pmatrix} z_{VV} & z_{V\rho} & z_{VT} \\ a_\rho z_{VV} & \lambda_{\rho\rho} + a_\rho z_{V\rho} & \lambda_{\rho T} + a_\rho z_{VT} \\ a_T z_{VV} & \lambda_{T\rho} + a_T z_{V\rho} & \lambda_{TT} + a_T z_{VT} \end{pmatrix}, \tag{90a}$$

or, in more explicit form,

$$\mathbf{Z} = \begin{pmatrix} -\xi P_L^* & -3\xi V_* kT_0 & -3\xi k N_* \\ \xi P_L^* \frac{\rho_*}{V_*} & \lambda_{\rho\rho} + 3\xi P_* & \lambda_{\rho T} + 3\xi k \rho_*^2 \\ \xi P_L^* \frac{P_*}{C_V^*} & \lambda_{T\rho} + 3\xi \frac{(kT_0)^2}{c_V} & \lambda_{TT} + 3\xi \frac{k}{c_V} P_* \end{pmatrix} \tag{90b}$$

The determinant of $\mathbf{Z}$ is

$$\det \mathbf{Z} = z_{VV} \det \mathbf{\Lambda} \tag{91}$$

The cubic equation for eigenvalues of the matrix $\mathbf{Z}$ has the following form:

$$\kappa^3 - (Sp\mathbf{Z})\kappa^2 + B\kappa - \det \mathbf{Z} = 0, \tag{92}$$

where

$$B = \Delta_{V\rho} + \Delta_{VT} + \Delta_{\rho T}, \quad \Delta_{V\rho} = \det \mathbf{Z}_{V\rho} = z_{VV} \lambda_{\rho\rho}, \quad \Delta_{VT} = \det \mathbf{Z}_{VT} = z_{VV} \lambda_{TT}, \quad \Delta_{\rho T} = \det \mathbf{Z}_{\rho T} \tag{93}$$

and

$$\mathbf{Z}_{V\rho} = \begin{pmatrix} z_{VV} & z_{V\rho} \\ a_\rho z_{VV} & \lambda_{\rho\rho} + a_\rho z_{V\rho} \end{pmatrix}, \; \mathbf{Z}_{VT} = \begin{pmatrix} z_{VV} & z_{VT} \\ a_T z_{VV} & \lambda_{TT} + a_T z_{VT} \end{pmatrix}, \; \mathbf{Z}_{\rho T} = \begin{pmatrix} \lambda_{\rho\rho} + a_\rho z_{V\rho} & \lambda_{\rho T} + a_\rho z_{VT} \\ \lambda_{T\rho} + a_T z_{V\rho} & \lambda_{TT} + a_T z_{VT} \end{pmatrix} \tag{94}$$

Employing the trigonometric form of solution of eq. (92), the desired root is selected as

$$\kappa_1 = \frac{1}{3}\left\{-2\sqrt{(Sp\mathbf{Z})^2 - 3B} \cos\left(\frac{\phi}{3} - \frac{\pi}{3}\right) + Sp\mathbf{Z}\right\}, \tag{95}$$

where

$$\cos\phi = -\frac{U}{2\sqrt{-(Y/3)^3}}, \quad U = -\frac{2}{27}(Sp\mathbf{Z})^3 + \frac{B Sp\mathbf{Z}}{3} - \det \mathbf{Z}, \quad Y = B - \frac{(Sp\mathbf{Z})^2}{3} \tag{96}$$

The calculation of $\kappa_1$ solves the problem of determining the stationary nucleation rate.

## IV. DISCUSSION. KINETIC LIMITS

### A. Kinetic region of nucleation

Let us consider a consequence of the obvious physical condition $\det \Lambda > 0$. From eq.'s (73) and (76), $\lambda_{\rho T} \lambda_{T \rho} = (k/c_V) \widetilde{q}^2 \lambda_{\rho\rho}^2$. Hence

$$\det \Lambda = \lambda_{\rho\rho} \lambda_{TT} - \lambda_{\rho T} \lambda_{T\rho} = \lambda_{\rho\rho}\left(\lambda_{TT} - \frac{k}{c_V}\widetilde{q}^2 \lambda_{\rho\rho}\right) > 0, \tag{97a}$$

i.e.

$$\lambda_{TT} > \frac{k}{c_V}\widetilde{q}^2 \lambda_{\rho\rho} \tag{97b}$$

Substituting eq.'s (79) for $\lambda_{TT}$ and (73) for $\lambda_{\rho\rho}$, we obtain

$$\alpha > \widetilde{q}^2 k \beta u_1 \rho_*, \quad u_1 = \sqrt{kT_0/2\pi m} \tag{97c}$$

Employing eq. (85) for $\alpha$, we find finally

$$\frac{R_*(T_0, P_0)\rho_*(T_0, R_*)}{\zeta(T_0)} < 1, \quad \zeta(T_0) \equiv \frac{\theta_0}{\widetilde{q}^2 k \beta u_1} \tag{98}$$

The quantity $\zeta$ weekly depending on temperature is a characteristic of the given liquid. The explicit dependences of $R_*$ and $\rho_*$ are indicated for further analysis of this inequality.

Condition (98) is not satisfied a fortiori near the equilibrium curve $P_\infty(T_0)$, since $R_* \to \infty$ and $\rho_* \to \rho_\infty(T_0) = P_\infty(T_0)/kT_0$ here. If the pressure $P_0$ decreases at a fixed value of $T_0$, then $R_*$ and $\rho_*$ decrease also. Consequently, this inequality begins to be satisfied starting from some value of pressure $\widetilde{P}_0(T_0)$.

The thermodynamic region of nucleation is the region under the equilibrium curve: $P_0 < P_\infty(T_0)$ (the metastable region). The inequality $P_0 < \widetilde{P}_0(T_0)$ narrows the thermodynamic region of nucleation. In other words, the nucleation of bubbles can occur only in the region $P_0 < \widetilde{P}_0(T_0)$, where $\widetilde{P}_0(T_0)$ is implicitly given by the equation

$$\frac{R_*(T_0, \widetilde{P}_0)\rho_*(T_0, R_*(T_0, \widetilde{P}_0))}{\zeta(T_0)} = 1 \tag{99a}$$

together with eq. (31) for $R_*(T_0, P_0)$. Combining both these equations, we obtain one more representation of eq. (99a):

$$\frac{\tilde{P}_0 R_*(T_0, \tilde{P}_0)}{kT_0 \zeta(T_0) - 2\sigma} = 1 \tag{99b}$$

Since the region $P_0 < \tilde{P}_0(T_0)$ is obtained from the kinetic condition $\det \Lambda > 0$, it can be called the kinetic region of nucleation. On the other hand, nucleation in the thermodynamic region does not occur near the equilibrium curve $P_\infty(T_0)$, since the critical radius is large here and the nucleation rate is practically equal to zero. The nucleation rate becomes appreciable at some distance from the equilibrium curve. If the curve $\tilde{P}_0(T_0)$ lies sufficiently close to the curve $P_\infty(T_0)$, i.e. in the region of zero nucleation rate, then condition (98) does not impose any restriction on the nucleation process. Otherwise, if the curve $\tilde{P}_0(T_0)$ lies sufficiently deep in the metastable region, condition (98) can forbid the nucleation process at the given $(T_0, P_0)$ even if it is allowed by thermodynamics (the work $W_*$ yields an acceptable value of the exponential function in eq. (86)).

Inequalities (97b) and (97c) show the limiting character of the thermal process in the nucleation of bubbles: the heat exchange between a bubble and ambient liquid has to be sufficiently fast. The quantities $R_*$ and $\rho_*$ have to be sufficiently small for that, according to eq.'s (79) and (85).

As one of the limiting cases, a non-volatile liquid, $P_\infty(T_0) \to 0$, is considered below. Hence, $P_* \to 0$ and $\rho_* \to 0$ also. For such liquids, condition (98) is always satisfied.

### B. Limits in the two-variable $(V, \rho)$-theory

Let us assume that the kinetic limit $\lambda_{TT} \to \infty$ occurs in our $(V, \rho, T)$-theory (it is considered below), so the theory becomes two-dimensional with the variables $\rho$ and $V$. In this case, the equation for eigenvalues has the form

$$\kappa^2 - (Sp\mathbf{Z}_{V\rho})\kappa + \det \mathbf{Z}_{V\rho} = 0, \tag{100}$$

from where

$$\kappa_1 = \frac{1}{2}\left\{Sp\mathbf{Z}_{V\rho} - \sqrt{(Sp\mathbf{Z}_{V\rho})^2 - 4\det \mathbf{Z}_{V\rho}}\right\}, \tag{101}$$

$$Sp\mathbf{Z}_{V\rho} = z_{VV} + \lambda_{\rho\rho} + a_\rho z_{V\rho} = -\xi P_L^* + \lambda_{\rho\rho} + 3\xi P_*$$

The following kinetic limits with respect to $\lambda_{\rho\rho}$ are possible.

**(1)** $\lambda_{\rho\rho} \gg \xi P_L^*, 3\xi P_*$

In this case,

$$(Sp\mathbf{Z}_{V\rho})^2 \gg |\det \mathbf{Z}_{V\rho}| = |z_{VV}|\lambda_{\rho\rho} \tag{102a}$$

and

$$Sp\mathbf{Z}_{V\rho} > 0, \tag{102b}$$

so eq. (101) becomes as

$$\kappa_1 = \frac{\det \mathbf{Z}_{V\rho}}{Sp\mathbf{Z}_{V\rho}} = \frac{z_{VV}\lambda_{\rho\rho}}{z_{VV} + \lambda_{\rho\rho} + 3\xi P_*}, \tag{103}$$

from where, in view of condition **(1)**,

$$\kappa_1 = z_{VV} = -\xi P_L^* = \kappa^{(1D)} \tag{104}$$

**(2)** $\lambda_{\rho\rho} \ll \xi P_L^*, 3\xi P_*;\ \ 3P_* - P_L^* > 0,\ (3P_* - P_L^*) \sim P_L^*, 3P_*,$

so $\lambda_{\rho\rho} \ll \xi(3P_* - P_L^*)$

Conditions (102a) and (102b) take place in this case also; hence eq. (103) is valid as before. Now it yields

$$\kappa_1 = -\lambda_{\rho\rho}\frac{P_L^*}{3P_* - P_L^*} \tag{105}$$

**(3)** $\lambda_{\rho\rho} \ll \xi P_L^*, 3\xi P_*;\ \ |3P_* - P_L^*| \ll P_L^*, 3P_*,$

so $\lambda_{\rho\rho} \sim \xi|3P_* - P_L^*|$ and $|Sp\mathbf{Z}_{V\rho}| = O(\lambda_{\rho\rho})$

In this case, we have $(Sp\mathbf{Z}_{V\rho})^2 \ll |\det \mathbf{Z}_{V\rho}|$ instead of condition (102a), and eq. (101) yields

$$\kappa_1 = -\sqrt{|\det \mathbf{Z}_{V\rho}|} = -\sqrt{\lambda_{\rho\rho}\xi P_L^*} \tag{106}$$

independently of the sign of $Sp\mathbf{Z}_{V\rho}$.

**(4)** $\lambda_{\rho\rho} \ll \xi P_L^*, 3\xi P_*;\ \ 3P_* - P_L^* < 0,\ |3P_* - P_L^*| \sim P_L^*, 3P_*,$

so $\lambda_{\rho\rho} \ll \xi|3P_* - P_L^*|$

In this case, condition (102a) is satisfied again, but $Sp\mathbf{Z}_{V\rho} < 0$. Eq. (101) yields

$$\kappa_1 = Sp\mathbf{Z}_{V\rho} = -\xi|3P_* - P_L^*| = -\xi|3P_0 + 2P_L^*|, \tag{107}$$

where the equilibrium condition $P_* = P_0 + P_L^*$ is employed. The inequality $3P_* - P_L^* < 0$ means that $P_0 < -(2/3)P_L^*$. At high negative pressures $P_0$, $|P_0| \gg P_*$, the equilibrium condition looks as $P_0 = -P_L^*$, so eq. (107) becomes as eq. (104) – the one-dimensional result again.

The case (**1**) corresponds to fast kinetics for the density $\rho$. In other words, the process of exchange by particles between a bubble and surrounding liquid (the evaporation-condensation process) is rapid. Any deviation of the density from its equilibrium value $\rho_*$ relaxes rapidly, so always $\rho = \rho_*$. Correspondingly, the theory becomes one-dimensional [13] which is confirmed by eq. (104). The latter is the result of the one-dimensional theory.

The cases (**2**)-(**4**) correspond to slow kinetics for $\rho$. The pressure $P_0$ changes from case (**2**) to case (**4**) from positive or moderately negative to large negative values. Accordingly, $Sp\mathbf{Z}_{V\rho}$ changes from positive in case (**2**) to negative in case (**4**) values. Eq.'s (105) and (106) show that the nucleation process in cases (**2**) and (**3**) is limited by the evaporation-condensation process which is the slowest in these cases; the quantity $\kappa_1$ is determined by the parameter $\lambda_{\rho\rho}$ of this process. Also these equations show the significance of a multivariable theory. Namely, if a one-dimensional theory with $\kappa_1 = z_{VV}$ is used when case (**2**) takes place, a strongly overestimated value of the nucleation rate is obtained.

On the other hand, the inertial motion of the liquid determining the rate of the bubble volume change, $\dot{V}$, is the slowest process in case (**1**). Accordingly, $\kappa_1$ is determined by the parameter $z_{VV}$ of this process, eq. (104), as limiting one in this case. So, the conclusion from these results is the nucleation rate is determined by the *slowest* kinetic process in a system.

As the negative pressure $P_0$ increases in absolute value, the limitation slackens: from linear with respect to $\lambda_{\rho\rho}$ in eq. (105) it becomes square root in eq. (106) and vanishes at all in eq. (107).

### C. Cavitation

Let us consider the limit of cavitation [2] in a non-volatile liquid, or the formation of cavities at high negative pressures $P_0$. As noted above, $P_* \to 0$ and $\rho_* \to 0$ in this case; hence, $Sp\mathbf{Z}_{V\rho} = z_{VV} + \lambda_{\rho\rho}$ and eq. (101) directly yields the one-dimensional result $\kappa_1 = z_{VV}$.

This one-dimensional result is obtained as the thermodynamic limit, $P_* \to 0$, within the kinetic theory. On the other hand, it can be also obtained as purely thermodynamic limit according to the general theory of Ref. [13]. From eq. (48), the element $h_{\rho\rho}$ of the matrix **H** is equal to $kT_0 V_* / \rho_*$, i.e. $(h_{\rho\rho}/kT_0) \to \infty$ at $\rho_* \to 0$. Accordingly, we have for the density variance: $\sqrt{\langle(\rho - \rho_*)^2\rangle} \to 0$. This means that a small value of density is equilibrium; the cavity in a non-volatile liquid is equilibrium with respect to vapor density. In other words, the probability of deviation from $\rho_*$ is negligibly small; always $\rho = \rho_*$, accordingly, the dimensionality of the theory decreases by unity and it becomes one-dimensional [13]. Recalling the expression for $\lambda_{TT}$, eq. (79), we see that $\lambda_{TT} \to \infty$ at $\rho_* \to 0$, i.e. the condition specified in the beginning of Section B is satisfied. In summary, the cavitation in a non-volatile liquid within the $(V, \rho, T)$- theory is the limiting process to the one-dimensional $(V)$-theory. This limiting process is thermodynamic in $\rho$ ($h_{\rho\rho}/kT_0 \to \infty$) and kinetic in $T$ ($\lambda_{TT} \to \infty$).

Equation $\kappa_1 = z_{VV}$ regarded as a limiting case of eq. (107) at $|P_0| \gg P_*$ can be interpreted as follows. At high negative pressures nucleation occurs as cavitation in a non-volatile liquid, despite the presence of vapor in a bubble, i.e. vapor does not affect the nucleation process [2]. Accordingly, the kinetic limitation vanishes in case (**4**) and a small value of $\lambda_{\rho\rho}$ is of no importance. The mentioned above one-dimensional result $\kappa_1 = z_{VV}$ directly obtained from eq. (101) in the case $P_* \to 0$ takes place at any $\lambda_{\rho\rho}$ value. So, thermodynamic conditions take precedence over kinetic ones in a nucleation process [13]. The conclusion about Ref. [2] following from the foregoing is the use of a one-dimensional theory therein is justified.

The two-variable $(V, T)$-theory can be considered by similar way. It is obtained from the $(V, \rho, T)$-theory, when either thermodynamic or kinetic limiting process with respect to $\rho$ takes place. In this case, we use the matrix $\mathbf{Z}_{VT}$ instead of $\mathbf{Z}_{V\rho}$ and $Sp\mathbf{Z}_{VT} = z_{VV} + \lambda_{TT} + a_T z_{VT} = -\xi P_L^* + \lambda_{TT} + 3\xi(k/c_V)P_*$. However, now the situation differs from the $(V, \rho)$-theory considered.

At first, let us assume that the thermodynamic limiting process in $\rho$, $\rho_* \to 0$, takes place. As is shown above, $\lambda_{TT} \to \infty$ in this case, i.e. case (**1**) with respect to $\lambda_{TT}$, $\lambda_{TT} \gg \xi P_L^*, 3\xi(k/c_V)P_*$ and, as a consequence, equation (104) for $\kappa_1$ take place. This is already discussed the case of cavitation in a non-volatile liquid. So, the cavitation is contained as a limiting case in both the $(V, \rho)$- and $(V, T)$-theories.

Hence, the cases of small values of $\lambda_{TT}$ (cases (**2**)-(**4**) considered above) could occur in the two-variable $(V, T)$-theory only in the kinetic limiting process with respect to $\rho$, $\lambda_{\rho\rho} \to \infty$. However, this case cannot be realized in view of condition (97b) which forbids the chain "$\lambda_{\rho\rho} \to \infty$, $\lambda_{TT} \to 0$". Though the elements $h_{\rho\rho}$ and $h_{TT}$ are "symmetric" with respect to the replacement $\rho_* \leftrightarrow T_0$, kinetic condition (97b) breaks this symmetry. Nevertheless, the case of small values of $\lambda_{TT}$ can take place together with small values of $\lambda_{\rho\rho}$; it is studied below within the framework of the $(V,\rho,T)$-theory.

So, the two-dimensional $(V,T)$-theory yields nothing new in addition to the $(V,\rho)$-theory. More precisely, it does not exist as such.

### D. Kinetic limits in the $(V,\rho,T)$-theory

The explicit form of the quantities $\Delta_{\rho T}$ and $B$, eq. (93), is

$$\Delta_{\rho T} = \det \mathbf{\Lambda} + (1+2\widetilde{q})(3\xi\frac{k}{c_V}P_*)\lambda_{\rho\rho} + 3\xi P_* \lambda_{TT} \tag{108a}$$

$$B = z_{VV}(\lambda_{\rho\rho} + \lambda_{TT}) + \Delta_{\rho T}, \tag{108b}$$

where $\det \mathbf{\Lambda}$ is given by eq. (97a).

At first, let us consider the kinetic limit $\lambda_{TT} \to \infty$ employed above and show that it indeed leads to the two-variable $(V,\rho)$-theory. Physically, this limit means

(**1**) $\lambda_{TT} \gg \xi P_L^*,\ 3\xi P_*,\ \lambda_{\rho\rho},\ \widetilde{q}^2 \lambda_{\rho\rho}$

Conditions (102) of the $(V,\rho)$-theory are generalized in the given case as

$$(Sp\mathbf{Z})^2 \gg |B|, \qquad Sp\mathbf{Z} > 0 \tag{109}$$

Denote

$$\beta_1 = \frac{B}{(Sp\mathbf{Z})^2}, \qquad \beta_2 = \frac{\det \mathbf{Z}}{(Sp\mathbf{Z})^3}; \qquad |\beta_1| \ll 1, \quad |\beta_2| \sim \beta_1^2 \tag{110}$$

We have for case (**1**)

$$Sp\mathbf{Z} = \lambda_{TT} + o(\lambda_{TT}), \quad \beta_1 = \frac{z_{VV} + \lambda_{\rho\rho} + 3\xi P_*}{\lambda_{TT}} = \frac{Sp\mathbf{Z}_{V\rho}}{\lambda_{TT}}, \quad \beta_2 = \frac{z_{VV}\lambda_{\rho\rho}}{\lambda_{TT}^2} = \frac{\det \mathbf{Z}_{V\rho}}{\lambda_{TT}^2} \tag{111}$$

The quantities $U$, $Y$ and $\cos\phi$, eq. (96), are transformed as follows

$$U = -\frac{2(Sp\mathbf{Z})^3}{27}\left\{1 - \frac{9}{2}\beta_1 + \frac{27}{2}\beta_2\right\},$$

$$Y = -\frac{(Sp\mathbf{Z})^2}{3}(1-3\beta_1),$$

$$\cos\phi = 1 - \frac{27}{8}\beta_1^2 + \frac{27}{2}\beta_2 \tag{112}$$

Using the identity

$$\cos\left(\frac{\phi}{3} - \frac{\pi}{3}\right) = \frac{1}{2}\cos\frac{\phi}{3} + \frac{\sqrt{3}}{2}\sin\frac{\phi}{3}$$

and smallness of $\phi$, we obtain

$$\cos\left(\frac{\phi}{3} - \frac{\pi}{3}\right) = \frac{1}{2}\left(1 + \frac{\phi}{\sqrt{3}} - \frac{\phi^2}{18}\right) \qquad (113)$$

Comparing the expansion $\cos\phi = 1 - \phi^2/2$ with eq. (112), we find

$$\phi = 3\sqrt{3}\sqrt{\frac{\beta_1^2}{4} - \beta_2} \qquad (114)$$

Expanding the root in eq. (95) in small parameter $\beta_1$ and substituting eq. (113) together with eq. (114), we obtain up to linear in $\beta_1$ terms

$$\kappa_1 = (Sp\mathbf{Z})\frac{\beta_1 - \sqrt{\beta_1^2 - 4\beta_2}}{2} = \frac{1}{2}\left\{Sp\mathbf{Z}_{V\rho} - \sqrt{(Sp\mathbf{Z}_{V\rho})^2 - 4\det\mathbf{Z}_{V\rho}}\right\} = \kappa_1^{(V,\rho)}, \qquad (115)$$

eq.'s (111) were used here.

So, eq. (101) of the two-variable $(V,\rho)$-theory is obtained.

**(2)** $\lambda_{TT}, \lambda_{\rho\rho} \ll \xi P_L^*, 3\xi P_*, \quad Sp\mathbf{Z} > 0$

In this case, conditions (109) are satisfied as before, so eq. (115) for $\kappa_1$ takes place in which

$$Sp\mathbf{Z} = z_{VV} + 3\xi\left(1 + \frac{k}{c_V}\right)P_* + o(|z_{VV}|, 3\xi P_*) \qquad (116)$$

and $\beta_i$ are given by eq. (110).

Now eq. (115) yields

$$\kappa_1 = (Sp\mathbf{Z})\frac{\beta_1 - \sqrt{\beta_1^2 + 4|\beta_2|}}{2} = \vartheta\frac{B}{Sp\mathbf{Z}}, \quad \vartheta = \frac{1}{2}\left\{1 \mp \sqrt{1 + \frac{4\det\mathbf{Z}\,Sp\mathbf{Z}}{B^2}}\right\}, \qquad (117)$$

where the upper and lower signs correspond to $B > 0$ and $B < 0$, respectively and

$$\frac{B}{Sp\mathbf{Z}} = \frac{-P_L^*(\lambda_{\rho\rho} + \lambda_{TT}) + 3(1 + 2\tilde{q})(k/c_V)P_*\lambda_{\rho\rho} + 3P_*\lambda_{TT}}{3(1 + k/c_V)P_* - P_L^*} \sim \lambda_{\rho\rho}, \lambda_{TT} \qquad (118)$$

If $|\det\mathbf{Z}|Sp\mathbf{Z}/B^2 \ll 1$, eq. (117) is simplified as follows:

$$\kappa_1 = \begin{cases} \dfrac{z_{VV}\det\mathbf{\Lambda}}{B}, & B > 0 \\ \dfrac{B}{Sp\mathbf{Z}}, & B < 0 \end{cases} \qquad (119)$$

Eq.'s (117) - (119) generalize eq. (105) of the $(V,\rho)$-theory.

**(3)** $\lambda_{TT}, \lambda_{\rho\rho} \ll \xi P_L^*, 3\xi P_*; \quad (\lambda_{TT} + \lambda_{\rho\rho}) \sim \xi\left|-P_L^* + 3(1 + k/c_V)P_*\right|, \quad |Sp\mathbf{Z}| = O(\lambda_{TT} + \lambda_{\rho\rho})$

In this case, $B < 0$ and $(Sp\mathbf{Z})^2 \ll |B|$. Denoting

$$\overline{\beta}_1 = \frac{(Sp\mathbf{Z})^2}{|B|} \ll 1, \quad \overline{\beta}_2 = 9\frac{Sp\mathbf{Z}}{|B|^{1/2}} + 27\frac{\det \mathbf{Z}}{|B|^{3/2}}, \quad \overline{\beta}_2 \sim \overline{\beta}_1^{1/2}, \qquad (120)$$

we find

$$U = -\frac{|B|^{3/2}(\overline{\beta}_2 + 2\overline{\beta}_1^{3/2})}{27}$$

$$Y = B\left(1 + \frac{1}{3}\overline{\beta}_1\right)$$

$$\cos\phi = \frac{1}{6\sqrt{3}}\left(\overline{\beta}_2 + O(\overline{\beta}_1^{3/2})\right) \qquad (121)$$

The last equation implies $\phi = \pi/2 - \phi'$, where $\phi'$ is a small quantity, hence

$$\cos\left(\frac{\phi}{3} - \frac{\pi}{3}\right) = \frac{\sqrt{3}}{2}\left(1 - \frac{\phi'}{3\sqrt{3}} - O(\phi'^2)\right) \qquad (122)$$

On the other hand, $\cos\phi = \sin\phi' = \phi'$, so $\phi'$ is determined by eq. (121). Expanding the root in eq. (95) in small parameter $\overline{\beta}_1$ and substituting eq. (122), we find

$$\kappa_1 = -\sqrt{|B|}\left\{1 + O(\overline{\beta}_1^{1/2})\right\} \qquad (123)$$

This equation generalizes eq. (106) of the $(V, \rho)$-theory.

**(4)** $\lambda_{TT}, \lambda_{\rho\rho} \ll \xi P_L^*, 3\xi P_*; \quad (Sp\mathbf{Z})^2 \gg |B|$ and $Sp\mathbf{Z} < 0$

We employ the same designations $\beta_1$ and $\beta_2$ as in case **(1)**; now $\beta_2 > 0$. Instead of eq. (112), we have

$$\cos\phi = -\left\{1 - \frac{27}{8}\beta_1^2 + \frac{27}{2}\beta_2\right\} \qquad (124)$$

From here, $\phi = \pi - \phi'$, $\phi'$ is small, and $\cos\phi = -(1 - \phi'^2/2)$. Comparing with eq. (124), we find $\phi'$ and

$$\cos\left(\frac{\phi}{3} - \frac{\pi}{3}\right) = \cos\frac{\phi'}{3} = 1 - \frac{3}{2}\left(\frac{\beta_1^2}{4} - \beta_2\right) \qquad (125)$$

Expanding the root in eq. (95) in $\beta_1$ with regard for the condition $Sp\mathbf{Z} < 0$ as well as eq. (125), we find

$$\kappa_1 = Sp\mathbf{Z}\left\{1 - \beta_1 + O(\beta_1^2)\right\} = -\xi\left|3(1 + k/c_V)P_* - P_L^*\right| \qquad (126)$$

This equation generalizes eq. (107) of the $(V, \rho)$-theory. At high negative pressures $P_0$, $|P_0| \gg P_*$, it also yields the one-dimensional eq. (104).

So, the same physical picture as in the two-variable $(V, \rho)$-theory takes place here, when the pressure $P_0$ changes from positive to high negative values in cases (**2**)-(**4**). At a positive or moderately negative pressure, the nucleation process is limited by slow kinetics of both kinds (the exchange by particles and the heat exchange), eq.'s (117) and (123). As the negative pressure increases in absolute value, the limitation slackens, eq. (123), and vanishes at high values of pressure, eq. (126).